\documentclass{iaus}
\usepackage{graphicx}

\title[~~Dust effects on the derived S{\'e}rsic indexes]
{Dust effects on the derived S{\'e}rsic indexes of disks and bulges in spiral galaxies}

\author[Bogdan A. Pastrav et al.]
{Bogdan A. Pastrav$^1$, Cristina C. Popescu$^1$, Richard J. Tuffs$^2$ \and Anne E. Sansom$^1$}
\affiliation{$^1$Jeremiah Horrocks Institute, University of Central Lancashire, Preston PR1 2HE, UK \\email: {\tt bapastrav@uclan.ac.uk} \\[\affilskip]

$^2$Max Planck Institut f{\"ur} Kernphysik, Saupfercheckweg 1, D-69117 Heidelberg, Germany}

\pubyear{2011} 
\volume{284}  
\pagerange{1--12}
\jname{The Spectral Energy Distribution of Galaxies}
\editors{R.J. Tuffs \&  C.C.Popescu, eds.}

\begin{document}

\maketitle

\begin{abstract}

We present a theoretical study that quantifies the effect of dust
on the derived S{\'e}rsic indexes of 
disks and bulges. The changes in the derived parameters from their 
intrinsic values (as seen in the absence of dust) were obtained by 
fitting S{\'e}rsic distributions on simulated images of disks and bulges 
produced using radiative transfer calculations and the model of 
\cite[Popescu et al. 2011]{Pop11}. We found that dust
has the effect of lowering the measured S{\'e}rsic index in most cases,
with stronger effects for disks and bulges seen through more optically
thick lines of sight.


\end{abstract}

\firstsection

\section{Introduction}

In recent years deep wide field spectroscopic and photometric 
surveys of galaxies (e.g. GAMA, \cite[Driver et al. 2011]{Dri11}) are providing us 
with large statistical samples of galaxies with good quality imaging
up to z=0.1. In parallel, automatic routines like GALFIT (\cite[Peng et
al. 2002]{Peng02}, \cite[Peng et al. 2010]{Peng10}) or GIM2D (\cite[Simard et al. 2002]{Sim02}) have been developed to address
the need of fitting large number of images of galaxies with 2D
analytic  functions to characterise the surface brightness
distribution of their stellar components. In particular S{\'e}rsic
functions are the most common distributions that are used to describe 
the profiles of galaxies and their constituent morphological
components. The derived S{\'e}rsic indexes are then used to classify
galaxies as disk or bulge dominated ones (e.g. 
\cite[Lee et al. 2012]{Lee12}) or in terms of a bulge-to-disk ratio when bulge/disk
decomposition is performed (\cite[Simard et al. 2011]{Sim11}).

One potential problem with the interpretation of the results of S{\'e}rsic
fits is that the measured S{\'e}rsic parameters differ from the intrinsic
ones (as would be derived in the absence of dust). This is because 
 real galaxies, in particular spiral galaxies contain large amount
 of dust, and this dust changes their appearance  from what
would be predicted to be seen in projection based on only their
intrinsic stellar distributions. Determining the changes due to dust
is thus essential when characterising and classifying galaxies based
on their fitted S{\'e}rsic indexes.

Here we present results of a theoretical study to quantify the effects
of dust on the measured fitted S{\'e}rsic indexes, as a function of the
relevant parameters: dust opacity, disk
inclination and wavelength. The results are presented for pure disks and
bulges, as seen through a common distribution of dust (in the disk of
the parent galaxy). These results are part of a larger study that aims
to quantify the effects of dust on all photometric parameters of young
stellar disks, old stellar disks and bulges (Pastrav et al., in prep)
and builds on our previous work of \cite[M\"ollenhoff et al. 2006]{Mol06}.

\section{The method}

We based our study on simulated images of dust-attenuated disks and 
bulges for which their intrinsic parameters are known, being input for
the simulations. We then measured the perceived (apparent) 
photometric parameters using widely used 2D fitting routines, in an 
attempt to exactly mimic what an observer would measure for a real 
galaxy. The difference between the measured parameters 
(affected by a combination of projection and dust effects) and the 
intrinsic ones (as used to construct the 3D distributions of stellar 
emissivity) provides knowledge on the effects of dust
(and also on projection effects). The aim of the study is thus to
provide observers with corrections for the ``simplicity'' of the
templates commonly used to analyse images of galaxies, ``simplicity''
which is nevertheless necessary for practical purposes when dealing
with large numbers of objects. 

The simulated images were
produced using radiative transfer calculations and the model of 
\cite[Popescu et al. 2011]{Pop11} (see also \cite[Popescu et al. 2000]{Pop00} and
\cite[Tuffs et al. 2004]{Tuf04}). In brief the disks were built using a distribution
of stellar emissivity described by a double exponential (in radial and vertical direction) while bulges 
were built using the
deprojection of general S{\'e}rsic distributions, in particular of a
distribution with a S{\'e}rsic index $n_{0}^{sers}=4$ (de Vaucouleurs) and a
S{\'e}rsic index $n_{0}^{sers}=1$ (exponential). Both disks and bulges were
attenuated by a common distribution of dust (present only in the
disk). More details about the geometry of the model can be found in 
\cite[Popescu et al. (2011)]{Pop11}.

To fit the simulated images, we applied the commonly used GALFIT 
data analysis algorithm (\cite[Peng et al. 2002]{Peng02}, \cite[Peng
et al. 2010]{Peng10}). Both disks and bulges were fitted by the
commonly used 2D S{\'e}rsic distributions, corresponding to an infinitely
thin disk. In both cases the S{\'e}rsic index of the fitted function was
left as a free parameter of the fit.

\section{Results}

\begin{figure}[t]
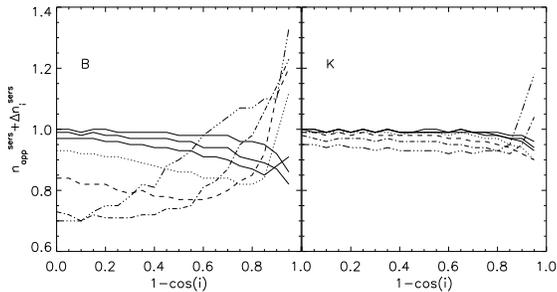

\hspace{2.5cm}
 \includegraphics[scale=0.325]{proc_var_orig_sersic_index_vs_inclination_disk_obs_b.epsi}
 \hspace*{-0.35cm}
 \includegraphics[scale=0.325]{proc_var_orig_sersic_index_vs_inclination_disk_obs_k.epsi}
\caption{The inclination dependence of the derived S\'{e}rsic index,
  $n_{app}^{sers}$,
  corrected for projection effects, $\Delta n_{i}^{sers}$, for disk
  images in the B and K bands. From top to bottom, the curves are
  plotted for central face-on opacity in the B band $\tau_{B}^{f}=0.1,0.3,0.5$ (solid lines), $1.0$ (dotted), 
$2.0$ (dashed), $4.0$ (dashed-dotted), $8.0$ (dashed-3 dotted).}
 \label{sersic_index_disk}
\end{figure}

\begin{figure}[t]
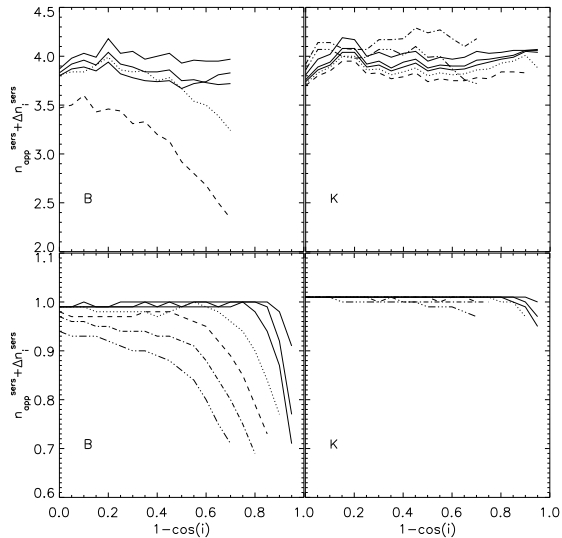

\hspace{2.75cm}
\includegraphics[scale=0.325]{proc_sersic_index_vs_inclination_corr_proj_bulge_obs_b.epsi}
\hspace{-0.21cm}
\includegraphics[scale=0.325]{proc_sersic_index_vs_inclination_corr_proj_bulge_obs_k.epsi}

\vspace{-0.03cm}

\hspace{2.75cm}
\includegraphics[scale=0.325]{proc_sersic1_index_vs_inclination_corr_proj_bulge_obs_b.epsi}
\hspace{-0.34cm}
\includegraphics[scale=0.325]{proc_sersic1_index_vs_inclination_corr_proj_bulge_obs_k.epsi}
\caption{Upper row: The inclination dependence of the derived
  S\'{e}rsic index, $n_{app}^{sers}$, 
  corrected for projection effects, $\Delta n_{i}^{sers}$, for de
  Vaucouleurs bulges ($n_{0}^{sers}=4$), in the B and K optical bands. 
  Lower row: The same, but for exponential bulges ($n_{0}^{sers}=1$). Line styles are as in Fig.~1.} 
\label{sersic_index_bulge}
\end{figure}
The results on the effects of dust on the derived S{\'e}rsic indexes are
summarised in Fig.~\ref{sersic_index_disk} (for the disk) and
Fig.~\ref{sersic_index_bulge} (for the bulge). In each figure we plot
the results on the measured S{\'e}rsic index corrected for projection
effects. By projection effects we mean the difference between the
projection of a distribution of stellar emissivity that has a vertical
distribution in addition to a radial distribution, and the
projection of an infinitely thin disk, which is assumed when fitting
the images with analytical 2D S{\'e}rsic functions. This results in an
alteration of the measured S{\'e}rsic index, even in the absence of
dust. The corrections for projection effects are discussed and listed
in Pastrav et al. (in prep). Here we only show the results on the
measured S{\'e}rsic index after correction for projection effects. 

In each figure we show the results as a function of inclination and
for different values of central face-on opacity in the B band,
$\tau_B^f$, which are the two parameters of the radiative transfer model. 
In addition we 
show results at two different wavebands (B and K). 

One can see that at longer wavelengths, in the K band, the effects
of dust are negligible. Essentially one can measure the intrinsic
S{\'e}rsic indexes for both disks and bulges, once corrected for 
projection
effects. For example for intrinsically exponential disks the derived 
S{\'e}rsic index is $\sim 1$. Similarly, for intrinsically 
de Vaucouleurs and exponential bulges the derived
S{\'e}rsic indexes are $\sim 4$ and $\sim 1$,
respectively. At shorter wavelength, in the B band, the dust starts to
affect the derived S{\'e}rsic indexes, with stronger effects for higher
$\tau^f_B$ and larger inclinations. In most cases the effect of 
dust is {\it to lower the S{\'e}rsic index} from its intrinsic value. This is
because of the large scale distribution of dust, which decreases
exponentially with increasing radial distance, making the profiles
flatter in the central parts.

Another result is that in most cases the {\it S{\'e}rsic
index decreases with increasing inclination and $\tau^f_B$}, except for
very optically thick disks,  which show a reverse trend.

\end{document}